# Photoionization of Xe 3$d$ electrons in molecule Xe@C$_{60}$: interplay of intra-doublet and confinement resonances

M. Ya. Amusia$^{1, 2}$, A. S. Baltenkov$^{3}$ and L. V. Chernysheva$^{2}$

$^{1}$Racah Institute of Physics, The Hebrew University, Jerusalem 91904, Israel
$^{2}$Ioffe Physical-Technical Institute, St.-Petersburg 194021, Russia
$^{3}$Arifov Institute of Electronics, Tashkent, 700125, Uzbekistan

**Abstract**

We demonstrate rather interesting manifestations of co-existence of resonance features in characteristics of the photoionization of 3$d$-electrons in Xe@C$_{60}$. It is shown that the reflection of photoelectrons produced by the 3$d$ Xe photoionization affects greatly partial photoionization cross-sections of $3d_{5/2}$ and $3d_{3/2}$ levels and respective angular anisotropy parameters, both dipole and non-dipole adding to all of them additional maximums and minimums. The calculations are performed treating the 3/2 and 5/2 electrons as electrons of different kinds with their spins "up" and "down". The effect of C$_{60}$ shell is accounted for in the frame of the "orange" skin potential model.

## 1. Introduction

Recently, a great deal of attention was and still is concentrated on photoionization of endohedral atoms. It was demonstrated in a number of papers that the C$_{60}$ shell adds prominent resonance structure. Although the experimental investigation of A@C$_{60}$ photoionization seems to be very difficult at this moment, it will be inevitably studied in the future. This justifies the current efforts of the theorists that are predicting rather non-trivial effects waiting for verification.

The role of C$_{60}$ in A@C$_{60}$ photoionization is manifold. C$_{60}$ act as a spherical potential resonator that reflects the photoelectron wave coming from A atom, thus leading to interference and so-called confinement resonances. C$_{60}$ at some frequencies acts as a dynamical screen that is capable to suppress or enhance the incident electromagnetic radiation acting upon the doped atom A. The reflection and refraction of the photoelectron waves by the potential resonator is prominent up to 60 – 80 eV of the electron energy. Whereas the screening effects of the C$_{60}$ shell are particularly strong for incident radiation frequency $\omega^{*/}$ of about that of the C$_{60}$ Giant resonance, i.e. 20 – 22 eV, but is noticeable in a much broader region, from ionization threshold up to again 60 – 80 eV.

Of special interest is the interference between atomic and C$_{60}$ resonances. An impressive example of it is the photoionization of 3$d$ Xe@C$_{60}$ where along with new so-called intra-doublet resonance that is a result of strong action of $3d_{3/2}$ level upon $3d_{5/2}$ one there is a strong action of the C$_{60}$ potential upon the electron waves that are emitted from both, $3d_{3/2}$ and $3d_{5/2}$ levels, which leads to interference phenomenon.

To consider the role of this interference upon all characteristics of the 3$d$ Xe@C$_{60}$ photoionization is the aim of the present paper.

---

$^{*}$ / Atomic system of units is used in this paper

At photon energy $\omega$ above the 3d subshell ionization threshold $I_{3d}$ the dynamic polarizability of $C_{60}$ and the screening effects of $C_{60}$ shell are already small enough. Therefore, incident photons freely penetrate inside the $C_{60}$ cavity and interacts with electrons of the doped Xe atom. On the other hand, even 50 – 60 eV above $I_{3d}$ the photoelectrons from 3d Xe interact strongly enough with $C_{60}$ being scattered both elastically and inelastic by the fullerene shell.

In this paper we will concentrate on elastic scattering as a more powerful effect that prominently modifies the cross section of 3d Xe while it is located inside of $C_{60}$. But we will not discuss here the inelastic process – the *exchange Auger decay* of $3d_{3/2}$ vacancy – that proceeds by its transition into $3d_{5/2}$ with emission of an electron from $C_{60}$.

## 2. Main formulas

We will start with the problem of an isolated atom. The method to treat the result of mutual action of $3d_{5/2}$ and $3d_{3/2}$ electrons for isolated atom Xe was discussed as two semi-filled levels with five spin-*up*($\uparrow$) and five spin-*down*($\downarrow$) electrons each was presented for the first time in [1]. Then the Random Phase Approximation with Exchange (RPAE) equations for atoms with semi-filled shells (so-called Spin Polarized RPAE or SP RPAE) are solved, as described in e.g. [2]. For semi-filled shell atoms the following relation gives the differential in angle photoionization cross-section by non-polarized light, which is similar to that of closed shell atoms [3] (see also e.g. [4]):

$$\frac{d\sigma_{nl\uparrow\downarrow}(\omega)}{d\Omega} = \frac{\sigma_{nl\uparrow\downarrow}}{4\pi}[1 - \frac{\beta_{nl\uparrow\downarrow}(\omega)}{2}P_2(\cos\theta) + \kappa\gamma_{nl\uparrow\downarrow}(\omega)P_1(\cos\theta) + \kappa\eta_{nl\uparrow\downarrow}(\omega)P_3(\cos\theta)], \quad (1)$$

where $\kappa = \omega/c$, $P_{1,2,3}(\cos\theta)$ are the Legendre polynomials, $\theta$ is the angle between photon $\mathbf{\kappa}$ and photoelectron momenta $\mathbf{k}$, $\beta_{nl\uparrow\downarrow}(\omega)$ is the dipole, while $\gamma_{nl\uparrow\downarrow}(\omega)$ and $\eta_{nl\uparrow\downarrow}(\omega)$ are so-called non-dipole angular anisotropy parameters, where the arrows $\uparrow\downarrow$ relates the corresponding parameters to *up* and *down* electrons, respectively.

Since in experiment, usually the sources of linearly polarized radiation are used, instead of (1) another form of angular distribution is more convenient [5, 6]:

$$\frac{d\sigma_{nl\uparrow\downarrow}(\omega)}{d\Omega} = \frac{\sigma_{nl\uparrow\downarrow}(\omega)}{4\pi}\{1 + \beta_{nl\uparrow\downarrow}(\omega)P_2(\cos\vartheta) + [\delta^C_{nl\uparrow\downarrow}(\omega) + \gamma^C_{nl\uparrow\downarrow}(\omega)\cos^2\vartheta]\sin\vartheta\cos\Phi\}. \quad (2)$$

Here $\vartheta$ is the polar angle between the vectors of photoelectron's velocity $\mathbf{v}$ and photon's polarization $\mathbf{e}$, while $\Phi$ is the azimuth angle determined by the projection of $\mathbf{v}$ in the plane orthogonal to $\mathbf{e}$ that includes the vector of photon's velocity. The non-dipole parameters in (1) and (2) are connected by the simple relations [4]

$$\gamma^C_{nl\uparrow\downarrow}/5 + \delta^C_{nl\uparrow\downarrow} = \kappa\gamma_{nl\uparrow\downarrow}, \qquad \gamma^C_{nl\uparrow\downarrow}/5 = -\kappa\eta_{nl\uparrow\downarrow}. \quad (3)$$

The below-presented results of calculations of non-dipole parameters are obtained using both expressions (1) and (2). There are two possible dipole transitions from subshell $l$, namely $l \to l \pm 1$ and three quadrupole transitions $l \to l; l \pm 2$. Corresponding general expressions for $\beta_{nl\uparrow\downarrow}(\omega)$, $\gamma_{nl\uparrow\downarrow}(\omega)$ and $\eta_{nl\uparrow\downarrow}(\omega)$ are rather complex and expressed via the dipole $d_{l\pm1}$ and quadrupole $q_{l\pm2,0}$ matrix elements of photoelectron transitions. In the one-electron Hartree-Fock (HF) approximation these parameters can be presented as [4, 7]:

$$\beta_{nl\uparrow\downarrow}(\omega) = \frac{1}{(2l+1)\left[(l+1)d_{l+1\uparrow\downarrow}^2 + ld_{l-1\uparrow\downarrow}^2\right]}[(l+1)(l+2)d_{l+1\uparrow\downarrow}^2 + l(l-1)d_{l-1\uparrow\downarrow}^2 -$$
$$6l(l+1)d_{l+1\uparrow\downarrow}d_{l-1\uparrow\downarrow}\cos(\delta_{l+1}-\delta_{l-1})] \qquad (4)$$

It is implied that the indexes $\uparrow\downarrow$ are added similarly to the parameters $\gamma_{nl}(\omega)$, $\eta_{nl}(\omega)$ and matrix elements $d_{l\pm1}$, $q_{l,l\pm2}$ in (5) and (6):

$$\gamma_{nl}(\omega) = \frac{3}{5\left[ld_{l-1}^2 + (l+1)d_{l+1}^2\right]}\left\{\frac{l+1}{2l+3}[3(l+2)q_{l+2}d_{l+1}\cos(\delta_{l+2}-\delta_{l+1}) - lq_ld_{l+1}\times\right.$$
$$\left.\cos(\delta_{l+2}-\delta_{l+1})] - \frac{l}{2l+1}[3(l-1)q_{l-2}d_{l-1}\cos(\delta_{l-2}-\delta_{l-1}) - (l+1)q_ld_{l-1}\cos(\delta_l-\delta_{l-1})]\right\}, \qquad (5)$$

$$\eta_{nl}(\omega) = \frac{3}{5\left[ld_{l-1}^2 + (l+1)d_{l+1}^2\right]}\left\{\frac{(l+1)(l+2)}{(2l+1)(2l+3)}q_{l+2}[5ld_{l-1}\cos(\delta_{l+2}-\delta_{l-1}) -\right.$$
$$-(l+3)d_{l+1}\cos(\delta_{l+2}-\delta_{l-1})] - \frac{(l-1)l}{(2l+1)(2l+1)}q_{l-2}\times$$
$$\times[5(l+1)d_{l+1}\cos(\delta_{l-2}-\delta_{l+1}) - (l-2)d_{l-1}\cos(\delta_{l-2}-\delta_{l-1})] + \qquad (6)$$
$$+2\frac{l(l+1)}{(2l-1)(2l+3)}q_l[(l+2)d_{l+1}\cos(\delta_l-\delta_{l+1}) - (l-1)d_{l-1}\cos(\delta_l-\delta_{l-1})]\right\}.$$

Here $\delta_l(k)$ are the photoelectrons' scattering phases; the following relation gives the matrix elements $d_{l\pm1\uparrow\downarrow}$ in the so-called $r$-form

$$d_{l\pm1\uparrow\downarrow} \equiv \int_0^\infty P_{nl\uparrow\downarrow}(r)rP_{\varepsilon l\pm1\uparrow\downarrow}(r)dr, \qquad (7)$$

where $P_{nl\uparrow\downarrow}(r)$, $P_{\varepsilon l\pm1\uparrow\downarrow}(r)$ are the radial so-called Spin-Polarized Hartree-Fock (SP HF) [8] one-electron wave functions of the $nl$ discrete level and $\varepsilon l\pm1$ - in continuous spectrum, respectively. The following relation gives the quadrupole matrix elements

$$q_{l\pm2,0\uparrow\downarrow} \equiv \frac{1}{2}\int_0^\infty P_{nl\uparrow\downarrow}(r)r^2 P_{\varepsilon l\pm2,l\uparrow\downarrow}(r)dr. \qquad (8)$$

In order to take into account the Random Phase Approximation with Exchange (RPAE) [3, 7] multi-electron correlations, one has to perform the following substitutions in the expressions for $\beta_{nl\uparrow\downarrow}(\omega)$, $\gamma_{nl\uparrow\downarrow}(\omega)$ and $\eta_{nl\uparrow\downarrow}(\omega)$ [4]:

$$d_{l+1}d_{l-1}\cos(\delta_{l+1}-\delta_{l-1}) \to [(\operatorname{Re}D_{l+1}\operatorname{Re}D_{l-1} + \operatorname{Im}D_{l+1}\operatorname{Im}D_{l-1})\cos(\delta_{l+1}-\delta_{l-1}) -$$
$$- (\operatorname{Re}D_{l+1}\operatorname{Im}D_{l-1} - \operatorname{Im}D_{l+1}\operatorname{Re}Q_{l-1})\sin(\delta_{l+1}-\delta_{l-1})] \qquad (9)$$

$$d_{l\pm1}q_{l\pm2,0}\cos(\delta_{l\pm2,0}-\delta_{l\pm1}) \to [(\operatorname{Re}D_{l\pm1}\operatorname{Re}Q_{l\pm2,0}+\operatorname{Im}D_{l\pm1}\operatorname{Im}Q_{l\pm2,0})\cos(\delta_{l\pm2,0}-\delta_{l\pm1})-$$
$$-(\operatorname{Re}D_{l\pm1}\operatorname{Im}Q_{l\pm2,0}-\operatorname{Im}D_{l\pm1}\operatorname{Re}Q_{l\pm2,0})\sin(\delta_{l\pm2,0}-\delta_{l\pm1})], \quad (10)$$
$$d_{l\pm1}^2 \to \operatorname{Re}D_{l\pm1}^2+\operatorname{Im}D_{l\pm1}^2.$$

The following are the ordinary RPAE equation for the dipole matrix elements

$$\langle v_2|D(\omega)|v_1\rangle = \langle v_2|d|v_1\rangle + \sum_{v_3,v_4}\frac{\langle v_3|D(\omega)|v_4\rangle(n_{v_4}-n_{v_3})\langle v_4 v_2|U|v_3 v_1\rangle}{\varepsilon_{v_4}-\varepsilon_{v_3}+\omega+i\eta(1-2n_{v_3})}, \quad (11)$$

where

$$\langle v_1 v_2|\hat{U}|v_1' v_2'\rangle \equiv \langle v_1 v_2|\hat{V}|v_1' v_2'\rangle - \langle v_1 v_2|\hat{V}|v_2' v_1'\rangle. \quad (12)$$

Here $\hat{V} \equiv 1/|\vec{r}-\vec{r}'|$ and $v_i$ is the total set of quantum numbers that characterize a HF one-electron state on discrete (continuum) levels. That includes the principal quantum number (energy), angular momentum, its projection and the projection of the electron spin. The function $n_{v_i}$ (the so-called step-function) is equal to 1 for occupied and 0 for vacant states.

For semi-filled shells the RPAE equations are transformed into the following system of equations that can be presented in the matrix form:

$$(\hat{D}_\uparrow(\omega)\hat{D}_\downarrow(\omega)) = (\hat{d}_\uparrow(\omega)\hat{d}_\downarrow(\omega)) + (\hat{D}_\uparrow(\omega)\hat{D}_\downarrow(\omega)) \times \begin{pmatrix}\hat{\chi}_{\uparrow\uparrow} & 0 \\ 0 & \hat{\chi}_{\downarrow\downarrow}\end{pmatrix} \times \begin{pmatrix}\hat{U}_{\uparrow\uparrow} & \hat{V}_{\uparrow\downarrow} \\ \hat{V}_{\downarrow\uparrow} & \hat{U}_{\downarrow\downarrow}\end{pmatrix}. \quad (13)$$

The dipole matrix elements $D_{l\pm1}$ are obtained by solving the radial part of the RPAE equation (11). As to the quadrupole matrix elements $Q_{l\pm2,0}$, they are obtained by solving the radial part of the RPAE equation, similar to (11)

$$\langle v_2|Q(\omega)|v_1\rangle = \langle v_2|\hat{q}|v_1\rangle + \sum_{v_3,v_4}\frac{\langle v_3|Q(\omega)|v_4\rangle(n_{v_4}-n_{v_3})\langle v_4 v_2|U|v_3 v_1\rangle}{\varepsilon_{v_4}-\varepsilon_{v_3}+\omega+i\eta(1-2n_{v_3})}. \quad (14)$$

Here in $r$-form one has $\hat{q} = r^2 P_2(\cos\theta)$.

Equations (13, 14) are solved numerically using the procedure discussed at length in [7]. The generalization of (14) for semi-filled shells is similar to (13):

$$(\hat{Q}_\uparrow(\omega)\hat{Q}_\downarrow(\omega)) = (\hat{q}_\uparrow(\omega)\hat{q}_\downarrow(\omega)) + (\hat{Q}_\uparrow(\omega)\hat{Q}_\downarrow(\omega)) \times \begin{pmatrix}\hat{\chi}_{\uparrow\uparrow} & 0 \\ 0 & \hat{\chi}_{\downarrow\downarrow}\end{pmatrix} \times \begin{pmatrix}\hat{U}_{\uparrow\uparrow} & \hat{V}_{\uparrow\downarrow} \\ \hat{V}_{\downarrow\uparrow} & \hat{U}_{\downarrow\downarrow}\end{pmatrix} \quad (15)$$

where

$$\hat{\chi}(\omega) = \hat{1}/(\omega-\hat{H}_{ev}) - \hat{1}/(\omega+\hat{H}_{ev}). \quad (16)$$

In (16) $\hat{H}_{ev}$ is the electron – vacancy HF Hamiltonian. Equations (13) and (15) permit to treat $3d_{5/2}$ and $3d_{3/2}$ electrons, if corrected by adding multipliers 6/5 and 4/5 to the *up* and *down* states, respectively. These equations were solved in this paper numerically. The cross-sections and angular anisotropy parameters are calculated by using numerical procedures with the codes described in [8].

### 3. Effect of C$_{60}$ fullerene shell

Since the thickness of the C$_{60}$ shell $\Delta$ is much smaller than its radius $R$, for low-energy photoelectrons one can substitute the C$_{60}$ potential by a zero-thickness pseudo-potential (see [9-12] and references therein):

$$V(r) = -V_0 \delta(r - R). \tag{17}$$

The parameter $V_0$ is determined by the requirement that the binding energy of the extra electron in the negative ion $C_{60}^-$ is equal to its observable value. Addition of potential (17) to the atomic HF potential leads to a factor $F_l(k)$ in the photoionization amplitudes which depends only upon the photoelectron's linear $k$ and angular $l$ moments [9-12]:

$$F_l(k) = \cos\Delta_l(k)\left[1 - \tan\Delta_l(k)\frac{v_{kl}(R)}{u_{kl}(R)}\right], \tag{18}$$

where $\Delta_l(k)$ are the additional phase shifts due to the fullerene shell potential (17). They are expressed by the following formula:

$$\tan\Delta_l(k) = \frac{u_{kl}^2(R)}{u_{kl}(R)v_{kl}(R) + k/2V_0}. \tag{19}$$

In these formulas $u_{kl}(r)$ and $v_{kl}(r)$ are the regular and irregular solutions of the atomic HF equations for a photoelectron with momentum $k = \sqrt{2\varepsilon}$, where $\varepsilon$ is the photoelectron energy connected with the photon energy $\omega$ by the relation $\varepsilon = \omega - I_A$ with $I$ being the atom A ionization potential.

Using Eq. (18), one can obtain the following relation for the $D^{AC}$ and $Q^{AC}$ amplitudes of endohedral atom expressed via the respective values for isolated atom that correspond to $nl \to \varepsilon l'$ transitions:

$$D(Q)_{nl,kl'}^{AC}(\omega) = F_{l'}(\omega)^2 D(Q)_{nl,kl'}(\omega). \tag{20}$$

For the cross-sections one has

$$\sigma_{nl,kl'}^{AC}(\omega) = |F_{l'}(\omega)|^2 \sigma_{nl,kl'}^A(\omega). \tag{21}$$

With these amplitudes, using the expressions (4-6) and performing the substitution (9, 10) we obtain the cross-sections for Xe@$C_{60}$ and angular anisotropy parameters. While calculating the anisotropy parameters, the cosines of atomic phases differences $\cos(\delta_l - \delta_{l'})$ in formulas (4)-(6) are replaced by $\cos(\delta_l + \Delta_l - \delta_{l'} - \Delta_{l'})$.

## 4. Some details of calculations and their results

Naturally, the parameters of $C_{60}$ were chosen the same as in previous papers, e.g. in [9]: $R = 6.639$ and $V_0 = 0.443$. In Fig. 1 and Fig. 2 we present the results for partial cross-sections that correspond to the $3d \to \varepsilon f$ and $3d \to \varepsilon p$ transitions, respectively. The solid line presents the $3d_{5/2}$ while the dashed line stands for $3d_{3/2}$ cross-sections of the Xe isolated atom. It is seen that the action of $3d_{3/2}$ electrons leads to an extra maximum in the $3d_{5/2}$ cross-section. The partial cross-section corresponding to the $3d \to \varepsilon p$ transition is two orders less than the $3d \to \varepsilon f$ partial cross-section. Therefore, below while considering the effect of the fullerene shell on the photoeffect in Xe@$C_{60}$ we will concentrate only on the main $3d \to \varepsilon f$ electron transition. In Fig. 3 we present the photoionization cross-section of $3d_{3/2}$ electrons of Xe@$C_{60}$ that has prominent oscillations due to reflection of the $\varepsilon f$ photoelectron wave by the $C_{60}$ shell. Fig. 4 demonstrates the effect of $C_{60}$ upon the photoionization cross-section of $3d_{5/2}$ electrons. Note that the additional maximum appearing due to the $3d_{3/2}$ action is prominently altered by the $C_{60}$ action and a number of additional maximums are created.

The amplitude factor (18) is defined by the values of the photoelectron wave functions with the "up" and "down" spins at the point $r = R$. Since we deal with the completely filled 3d Xe subshell the role of the spin effects in behavior of the wave function is small. Hence the continuum wave functions for the "up" and "down" spins at this point are almost equal. So, the amplitude factors $F_l(k)$ for the 3/2 and 5/2 levels are similar to each other. However, since the background cross-sections for the isolated Xe atom are significantly different, the interference of 3/2 and 5/2 intra-doublet resonances and confinement resonances leads to essentially different cross-sections for Xe@$C_{60}$ compounds, which we see in Figs. 3 and 4.

Figures 5-8 demonstrate the noticeable modifications in the angular anisotropy parameters, both dipole $\beta(\omega)$ and non-dipole ($\gamma^C(\omega)$, $\delta^C(\omega)$ and their combination $\gamma^C(\omega) + 3\delta^C(\omega)$) for $3d_{3/2}$ electrons. The dashed line gives the data for the isolated Xe, while the solid lines give the same for the molecule Xe@$C_{60}$. Figures 9-12 give the same as Figures 5-8 with the same notations but for the $3d_{5/2}$ electrons.

Analyzing the Figures, we see that the presence of the $C_{60}$ shell enhances prominently the maximum in $\beta_{5/2}(\omega)$ due to the $3d_{3/2}$ electron action. In general, the respective non-dipole parameters for the 3/2 and 5/2 electrons are similar. Particularly strong is the deviation from the isolated atom values close to thresholds. Entirely, we see that the presence of the $C_{60}$ shell leads to prominent extra additional resonance structure in all the characteristics of the photoionization of Xe@$C_{60}$ $3d_{5/2,3/2}$ electrons.

**Acknowledgments**

MYaA is grateful for financial support to the Israeli Science Foundation, Grant 174/03 and the Hebrew University Intramural Funds. ASB expresses his gratitude to the Hebrew University for hospitality and for financial support by Uzbekistan National Foundation, Grant Ф-2-1-12.

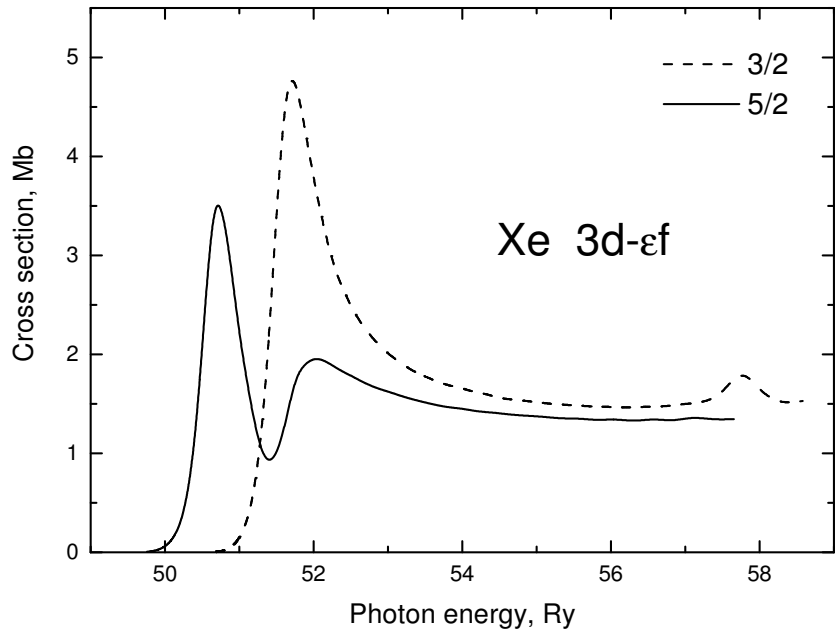

Fig.1. Photoionization cross-section of Xe 3d electrons, $3d - \varepsilon f$ transition

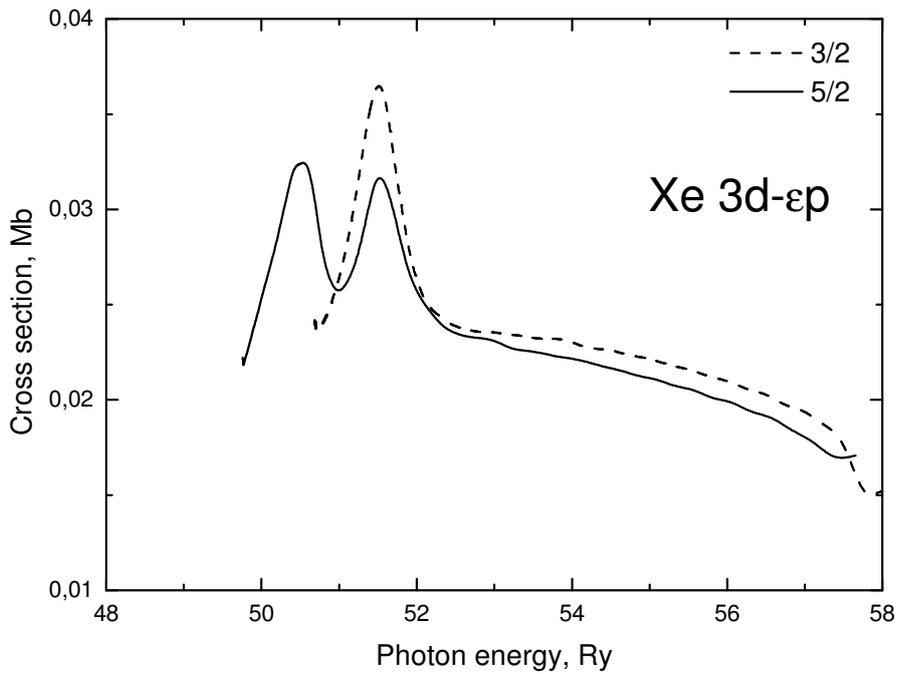

Fig. 2. Photoionization cross-section of Xe 3d electrons, $3d - \varepsilon p$ transition

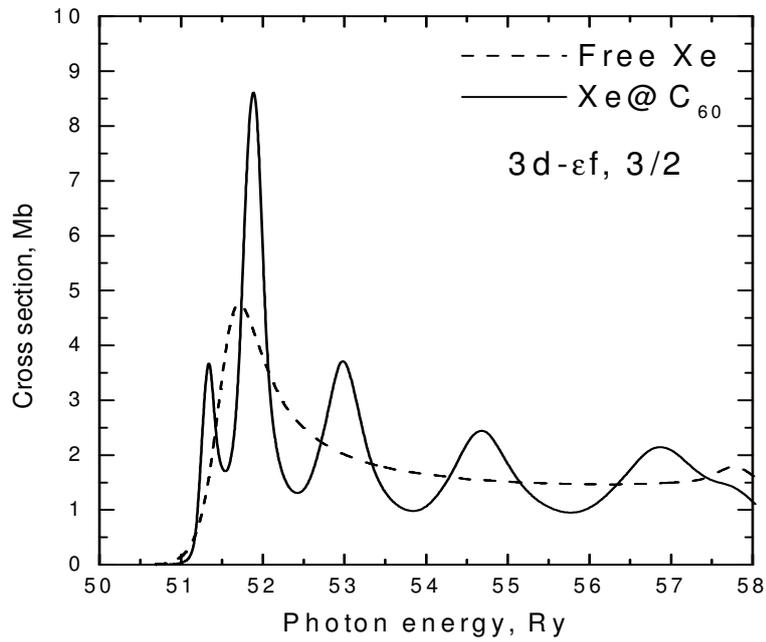

Fig. 3. Photoionization cross-section of $3d_{3/2}$ electrons in Xe and Xe@C$_{60}$.

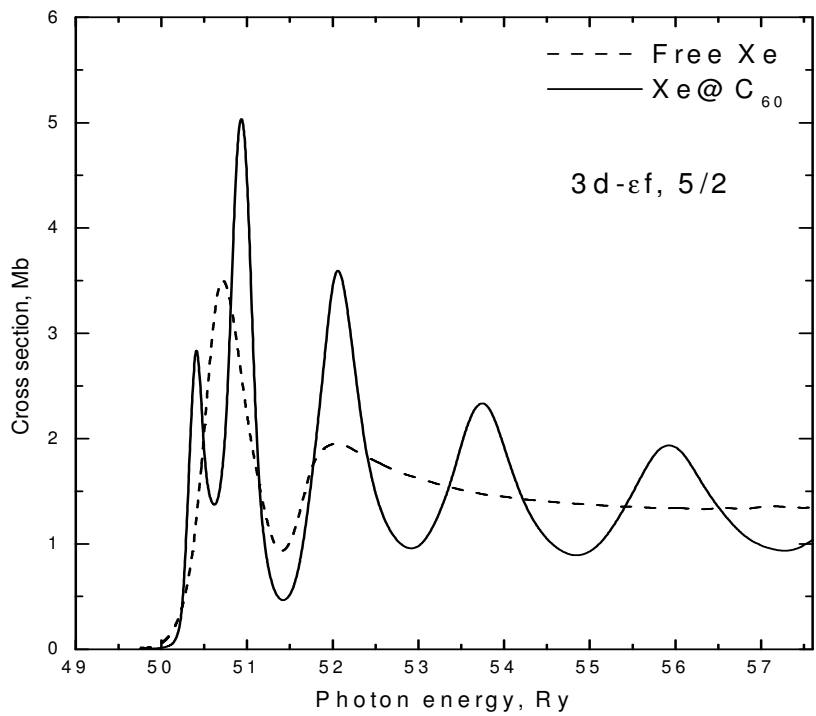

Fig. 4. Photoionization cross-section of $3d_{5/2}$ electrons in Xe and Xe@C$_{60}$.

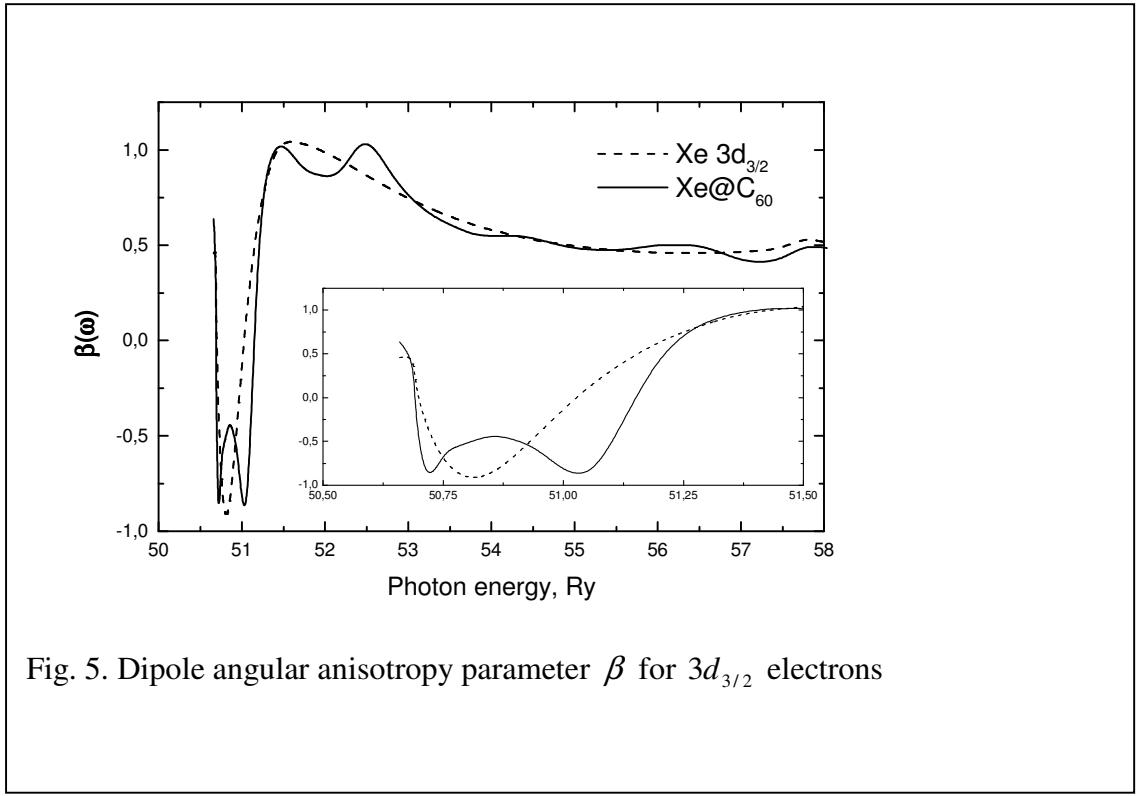

Fig. 5. Dipole angular anisotropy parameter $\beta$ for $3d_{3/2}$ electrons

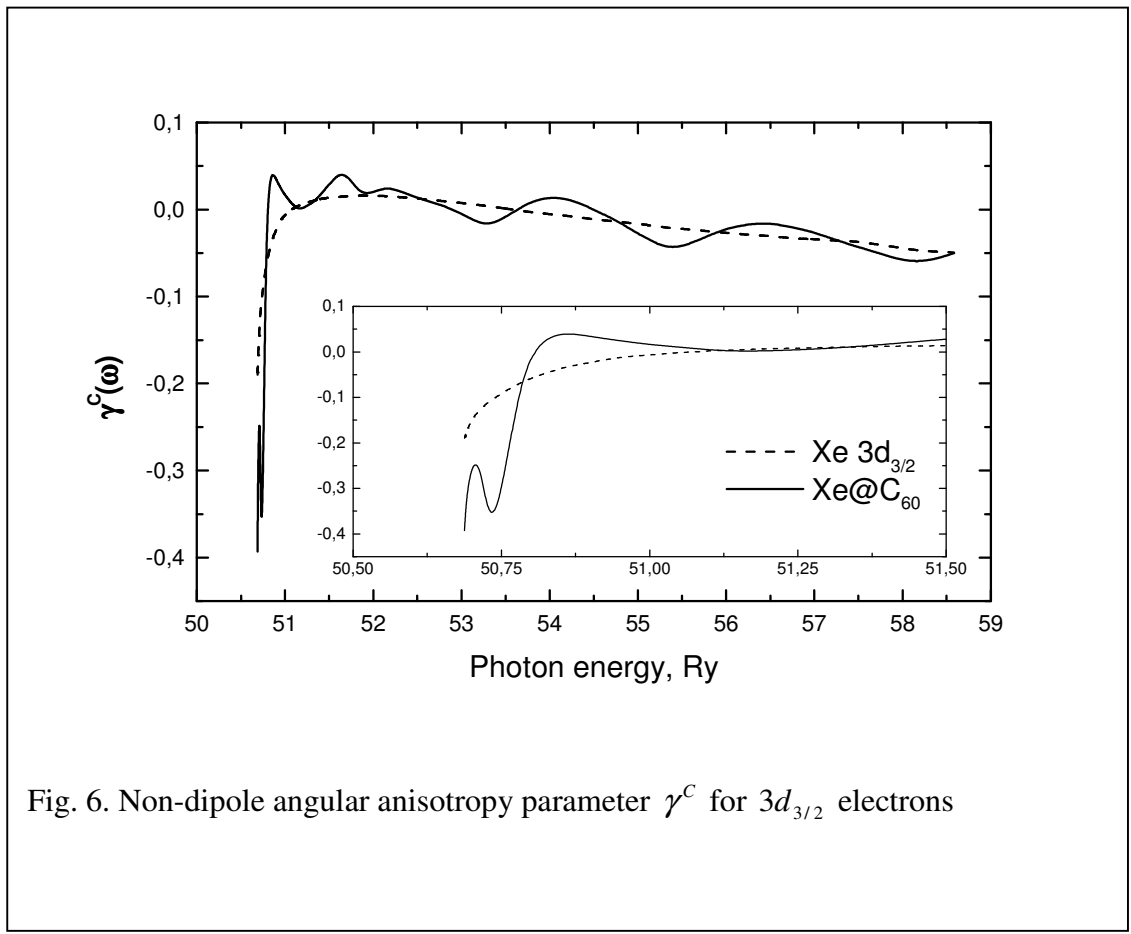

Fig. 6. Non-dipole angular anisotropy parameter $\gamma^C$ for $3d_{3/2}$ electrons

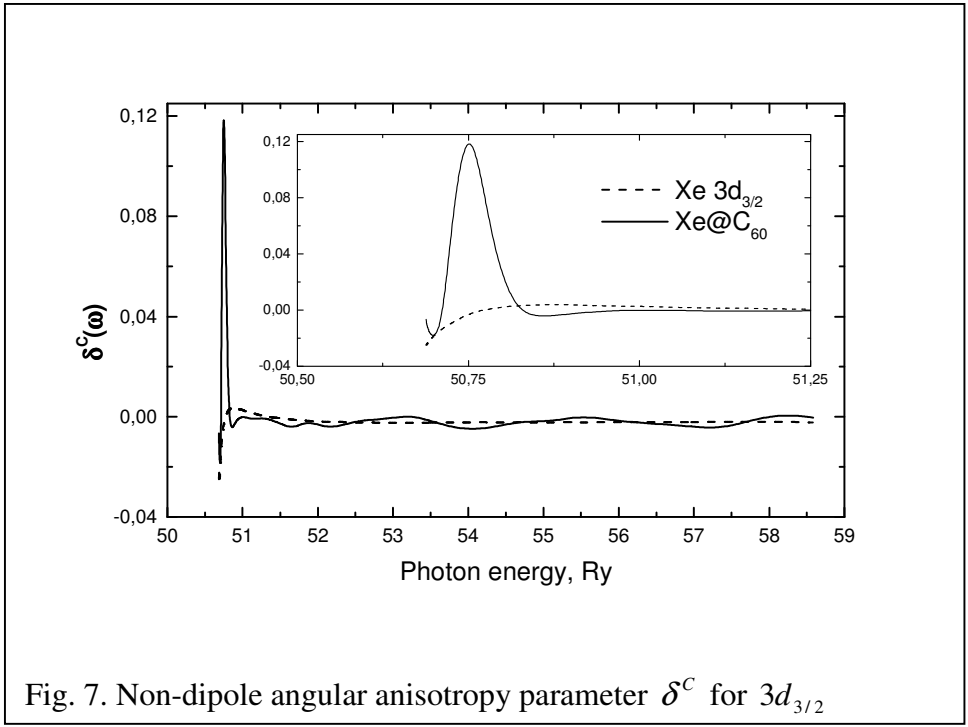

Fig. 7. Non-dipole angular anisotropy parameter $\delta^C$ for $3d_{3/2}$

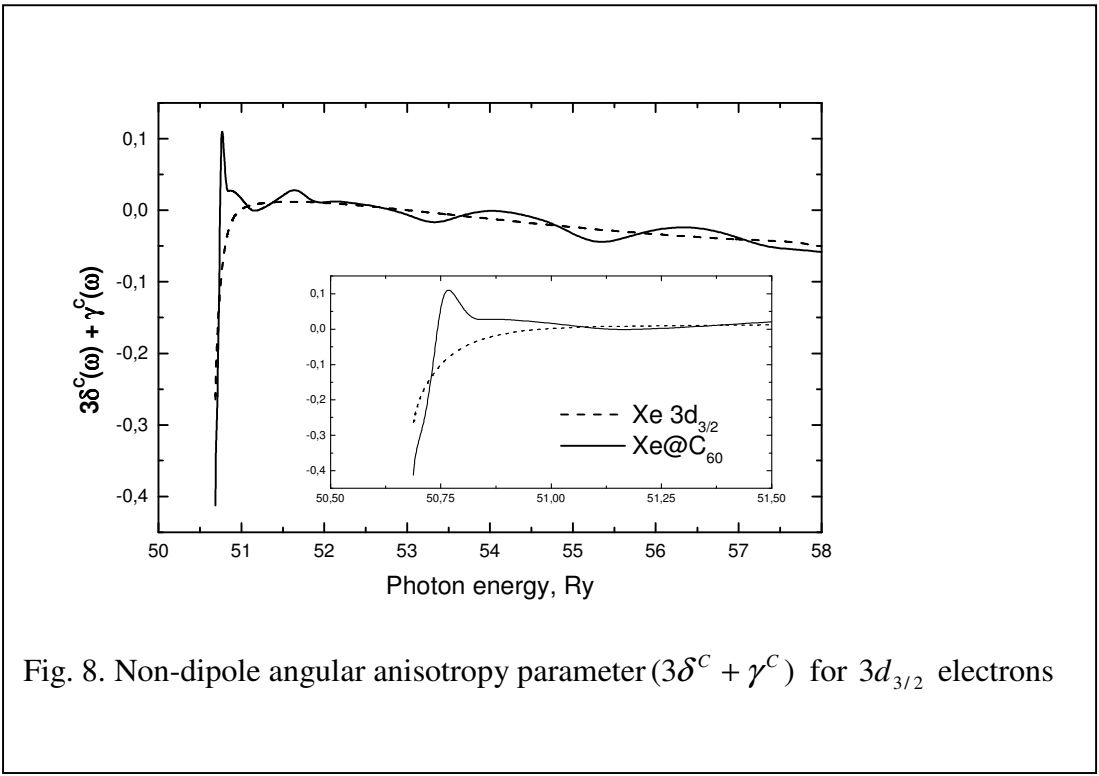

Fig. 8. Non-dipole angular anisotropy parameter $(3\delta^C + \gamma^C)$ for $3d_{3/2}$ electrons

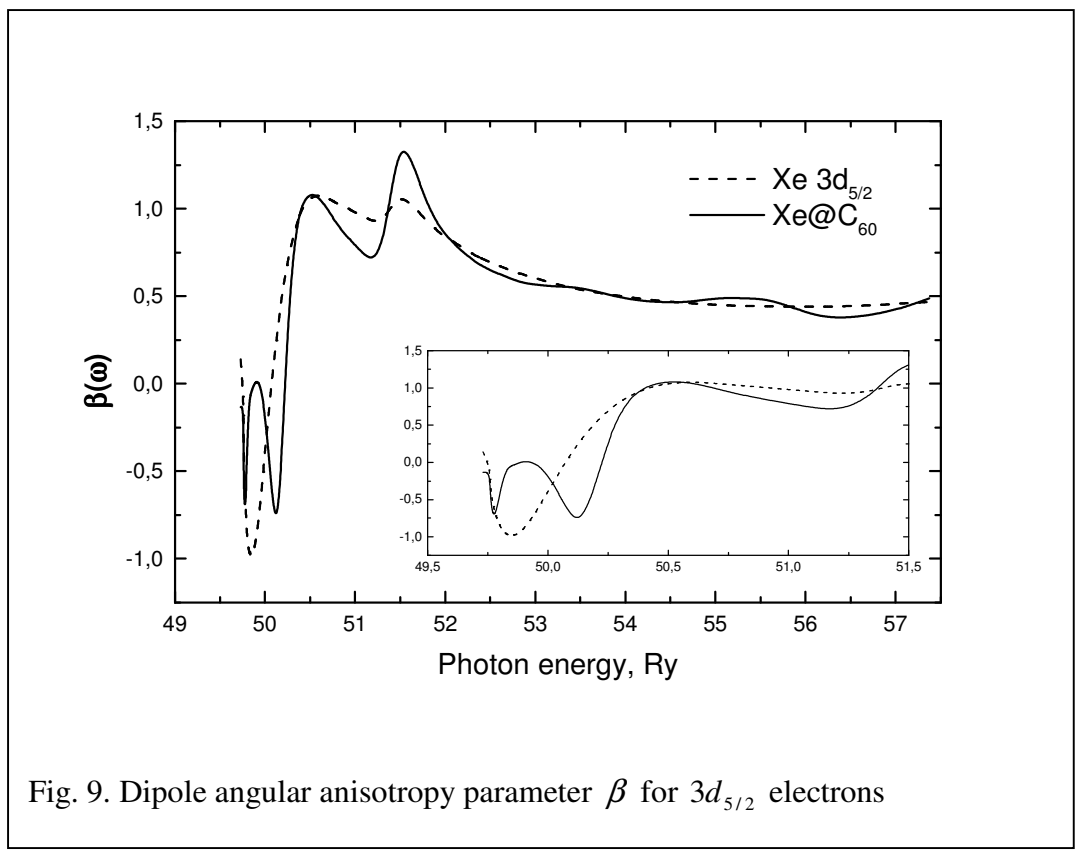

Fig. 9. Dipole angular anisotropy parameter $\beta$ for $3d_{5/2}$ electrons

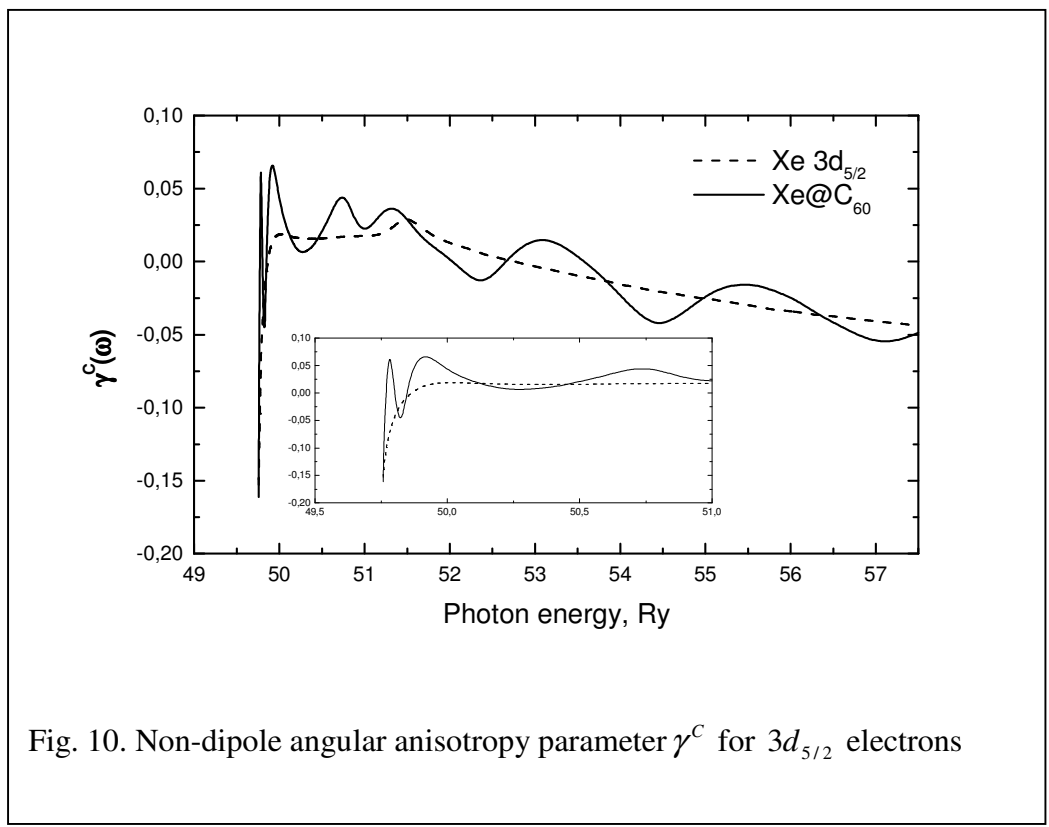

Fig. 10. Non-dipole angular anisotropy parameter $\gamma^C$ for $3d_{5/2}$ electrons

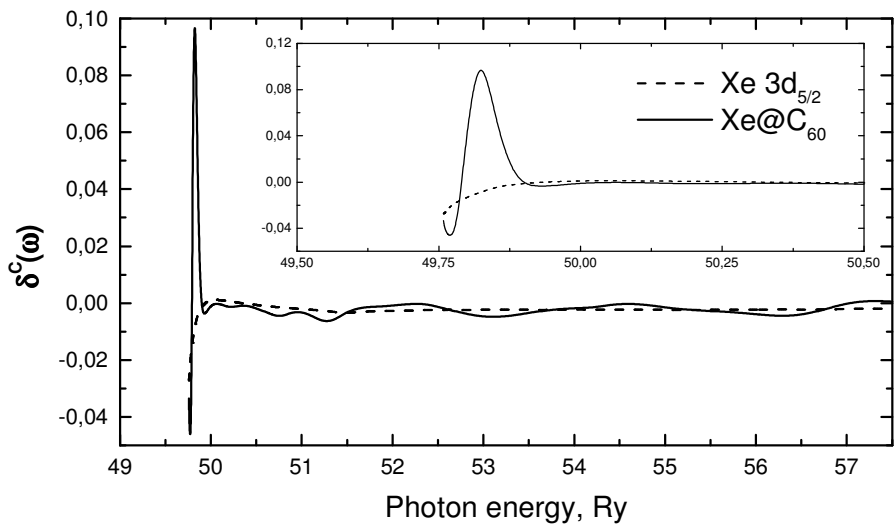

Fig. 11. Non-dipole angular anisotropy parameter $\delta^C$ for $3d_{3/2}$ electrons

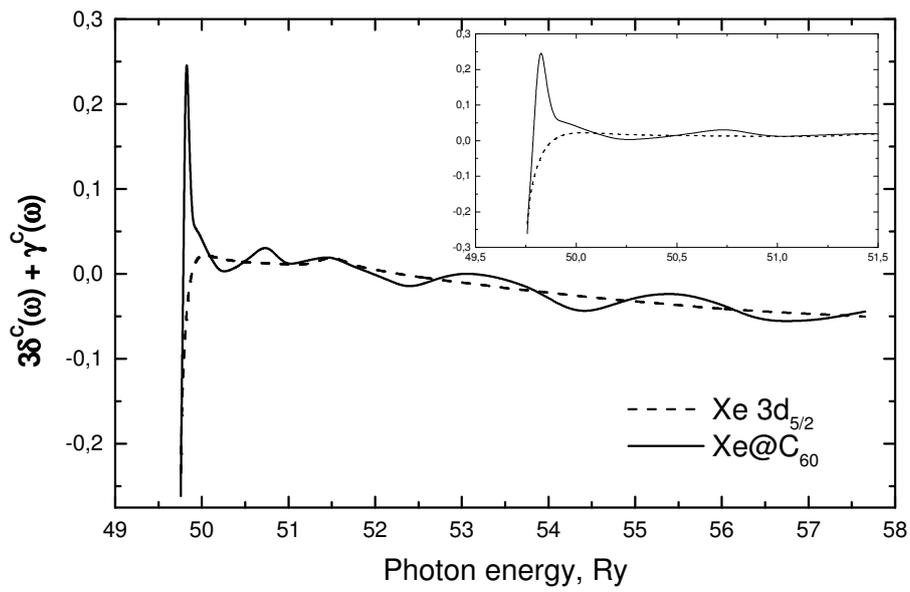

Fig. 12. Non-dipole angular anisotropy parameter $(3\delta^C + \gamma^C)$ for $3d_{5/2}$ electrons

13